# An Exchange-Based Diagnostic for Static Correlation

Jan M.L. Martin,[1, a)] Golokesh Santra,[1] and Emmanouil Semidalas[1]

[1]*Dept. of Molecular Chemistry and Materials Science, Weizmann Institute of Science, 7610001 Reḥovot, Israel*

a) Corresponding author: gershom@weizmann.ac.il

**Abstract.** We propose here a DFT-based diagnostic for static correlation %TAE$_X$[TPSS@HF – HF] which effectively measures how different the DFT and HF exchange energies for a given HF density are. This and %TAE$_{corr}$[TPSS] are two cost-effective a priori estimates for the adequacy of the importance of static correlation. %TAE$_X$[TPSS@HF – HF] contains nearly the same information as the earlier A diagnostic, but may be more intuitive to understand. Principal component and variable clustering analyses of a large number of static correlation diagnostics reveal that much of the variation is explained by just two components, and almost all of it by four; these are blocked by four variable clusters (single excitations; correlation entropy; double excitations; pragmatic energetics).

## INTRODUCTION

In wavefunction ab initio electronic structure theory, a molecule is said to be dominated by dynamical correlation if the wavefunction is dominated by a single reference determinant (although there may be minute contributions from billions of other Slater determinants), and hence that determinant is a good zero-order wave function in approximate treatments (such as by many-body perturbation theory,[1] MBPT). Static correlation, a.k.a. nondynamical correlation, refers to the situation where one or more additional determinants become prominent and hence a single Slater determinant becomes an ever poorer zero-order wave function. In such situations, single-reference MBPT will converge poorly with increasing order; coupled cluster methods are more resilient, and the widely used 'gold standard' CCSD(T) method[2,3] can handle mild to moderate static correlation owing to an error compensation.[4–7] Yet there is a definite need for reliable diagnostics for static correlation — preferably, inexpensive a priori estimates rather than expensive a posteriori procedures.

Fogueri et al.[8] have reviewed a number of common diagnostics (see below). In addition, they introduced a DFT-based diagnostic: if we define TAE as the total atomization energy of a molecule (i.e., the energy required to break up the molecule into its constituent ground-state atoms), then the slope of TAE with respect to the percentage of Hartree-Fock exchange in a hybrid functional, e.g., $A_{\lambda,PBE}$=(1-TAE[PBE$_\lambda$]/TAE[PBE])/$\lambda$ in the case of the widely used Perdew-Burke-Ernzerhof (PBE) functional,[9] turns out to be a surprisingly effective diagnostic.[8]

Handy and Cohen, in their paper introducing the OPTX ("optimized exchange") functional,[10] actually argue that DFT exchange can capture static correlation effects (at least the type A variety). They cite earlier work by Baerends and coworkers,[11] and indeed point to a 1953 paper by Slater[12] with a similar argument. In a nutshell: they considered the exchange energies of stretched diatomics like H$_2$, which have strong 'left-right correlation' (Handy and Cohen's term), 'type A static correlation' (Hollett and Gill's term,[13] also 'absolute near-degeneracy correlation'; "left-right strong correlation"[14]). In all cases, the DFT exchange energies are considerably more negative than their Hartree-Fock counterparts, by amounts comparable to what an MCSCF calculation involving the antibonding excited determinant would yield. As Handy and Cohen put it: "At infinite separation, H$_2$ has zero dynamic correlation. Thus, the non-dynamic correlation energy is an exchange integral. We deduce that *we cannot separate exchange and non-dynamic correlation*." [It appears that Hollett and Gill type B, or relative near-degeneracy correlation, or 'angular strong correlation',[14] is a different story: Handy and Cohen's OPTX functional reproduces the HF exchange energy of Be atom exactly,[10] while an MCSCF calculation that includes both the (2s)$^2$ and (2p)$^2$ configurations yields a considerably lower energy.]

It is comparatively straightforward to decompose computed DFT energies into 1-electron, nuclear repulsion, Coulomb, and exchange components: the sum of the first three adds up to a Hartree-like energy. One can then also decompose TAE in the same manner, and consider the difference 100%*(TAE$_{DFT}$ – TAE$_{DFT,Hartree}$)/TAE$_{DFT}$ = %TAE[X], where "X" represents the exchange energy component. Needless to say, however, %TAE[X] implies the somewhat "heretical" concept of a total atomization energy for a permutationally non-invariant density.

Somewhat less controversially, one could instead compare TAE[X,HF] with TAE[X,DFT] — or better still, TAE[X,HF] with TAE[X,DFTx@HF] where our notation refers to the evaluation of the DFT exchange energy from the Hartree-Fock density (a.k.a., DC-DFT, density-corrected DFT[15]), rather than self-consistently. When (for example, for the PBE functional) TAE[X,PBEx@HF] – TAE[X,HF] is small, it physically means that the HF exchange is very similar to semilocal exchange with the same density. When it is large, the HF exchange is qualitatively different from semilocal exchange.

We hence define a new diagnostic (using the Tao-Perdew-Scuseria-Staroverov, TPSS,[16] meta-GGA functional):
%TAE[ΔX,TPSS]=100%*(TAE[X,TPSS@HF] – TAE[X,HF])/TAE[CCSD(T)]
Let us now consider how it behaves when applied to a real-life sample. As an afterthought, we also included
%TAE[ΔX,KS-HF(TPSS)]=100%*(TAE[X,TPSS] – TAE[X,TPSS@HF])/TAE[CCSD(T)]

## METHODS

### Dataset selection

We start with the W4-17 dataset of Karton et al.[17] These are 200 small molecules for which we were able to obtain W4 or better thermochemical data, and for which we hence have coupled cluster components available as high up the ladder as CCSDTQ5. Correlation regimes range from the purely dynamical (CH$_4$, H$_2$O, n-pentane) to the pathologically static (O$_3$, BN($^1\Sigma^+$), C$_2$($^1\Sigma^+_g$),…)

As many diagnostics are ambiguously defined for open-shell systems, or behave differently for them, we restricted ourselves to the closed-shell subset. This leaves us with 160 molecules.

### Computational details

The costlier energy-based diagnostics were extracted from the supporting information (SI) of the W4-17 paper.[17] Remaining diagnostics were evaluated using MOLPRO 2020[18] with the cc-pVTZ basis set,[19] (or for the DFT-based diagnostics, using ORCA 5.[20] Geometries were used unchanged from the SI of Ref. [17]

Data analysis was carried out using the JMP Pro 16 statistical software suite (SAS, Inc.).

## RESULTS AND DISCUSSION

Aside from %TAE$_X$[TPSS@HF – HF], we included the following sixteen variables in our analysis: (a) the T$_1$ diagnostic of Lee and Taylor,[21] being the Euclidean norm of the single excitation amplitudes vector T$_1$=||**t**$_1$||$_2$/N$_{val}^{1/2}$ (division by the square root of the number of valence electrons N$_{val}$ ensures n identical systems have the same T$_1$ as a single system); (b) the D$_1$ diagnostic,[22] which is the square root of the largest eigenvalue of the matrix **t**$_1$.**t**$_1^T$ ; (c) its double excitations equivalent D$_2$;[23] (d,e) the largest single and double amplitudes in absolute value, max|t1| and max|t2|; (f) the logarithm of the sum of the CCSD cluster amplitudes divided by N$_{val}$ (again to ensure size-intensivity); (g) the Matito nondynamical correlation index[24,25] normalized by the total correlation index; (h) a variant revI$_{ND}$=2I$_{ND}$/N$_{val}$ normalized for size intensivity;[26] (i) –T$_{e,SCF}$, the negative first excitation energy at the SCF level (which will be positive for systems with an RHF-UHF instability); (j) the Truhlar M diagnostic, which for closed-shell molecules is obtained from the natural orbital occupation numbers of the HOMO and LUMO as M=(2 – n$_{HOMO}$ + n$_{LUMO}$)/2; (k) the normalized von Neumann correlation entropy[27–29] S$_{norm}$= S$_{vonNeumann}$/N$_{val}$, S$_{vonNeumann}$=∑$_i$ n$_i$ ln n$_i$ which is the indicator most identified with the concept of quantum entanglement;[29–31] (l) the percentage of connected triples in the total atomization energy %TAE[(T)] proposed in Ref.[32]; (m) the overall percentage of correlation energy in TAE[CCSD(T)]; (n) the same using the TPSS meta-GGA functional;[33] (o) the percentage accounted for by connected quadruples %TAE[T$_4$]; (p) A$_{100}$[TPSS].

Upon executing a variable clustering analysis, we find that four variable clusters exist. The said clusters are also clearly reflected in the structure of the eigenvectors of the correlation matrix, i.e., the principal components. Both are presented in Table 1.

We note that the eigenvalues of the first four principal components add up to 17.15 variables out of 18, and just the first two to 14.93 variables. Also, we note that the large ratio (3.82) between the 4th and 5th PC eigenvalues provides us with a convenient cutoff point, even larger than the 3.07 between the 2nd and 3rd eigenvalue.

**TABLE 1.** First four principal components of the correlation matrix for the W4-17closed dataset.

| Diagnostics | Eigenvalues | 10.6807 | 4.2522 | 1.3834 | 0.8308 | 0.2173 |
|---|---|---|---|---|---|---|
| $T_1$ |  | 0.245 | 0.072 | -0.454 | 0.127 | 0.001 |
| $D_1$ | Cluster 1 | 0.234 | -0.004 | -0.532 | -0.022 | 0.149 |
| $Max|T_1|$ |  | 0.228 | 0.026 | -0.544 | -0.013 | 0.001 |
| $2\ln(A)/N_{val}$ |  | 0.198 | 0.336 | 0.052 | 0.269 | -0.120 |
| $I_{ND}/(I_{ND}+I_D)$ | Cluster 2a | 0.237 | 0.264 | 0.110 | 0.159 | 0.039 |
| $revI_{ND,2021}$ |  | 0.166 | 0.365 | 0.126 | 0.310 | 0.036 |
| $S_{norm}$ |  | 0.138 | 0.386 | 0.226 | 0.260 | 0.198 |
| Truhlar M |  | 0.270 | 0.157 | 0.068 | -0.172 | -0.227 |
| $D_2$ | Cluster 2b | 0.253 | 0.206 | 0.114 | -0.273 | 0.057 |
| $Max|T_2|$ |  | 0.240 | 0.138 | 0.064 | -0.474 | -0.639 |
| $-T_{e,SCF}$ |  | 0.227 | 0.134 | 0.106 | -0.558 | 0.630 |
| %TAE[(T)] |  | 0.271 | -0.197 | 0.018 | 0.158 | 0.000 |
| %TAE[$T_4$] | Cluster 3 | 0.275 | -0.152 | 0.078 | 0.140 | -0.199 |
| %TAE$_{corr}$[CCSD(T)] |  | 0.249 | -0.264 | 0.120 | 0.115 | 0.083 |
| %TAE$_{corr}$[TPSS] |  | 0.249 | -0.261 | 0.113 | 0.118 | 0.111 |
| $A_{100}$[TPSS] |  | 0.247 | -0.274 | 0.083 | 0.069 | -0.001 |
| %TAE$_X$[TPSS@HF - HF] |  | 0.237 | -0.287 | 0.137 | 0.043 | -0.002 |
| %TAE$_X$[TPSS – TPSS@HF] |  | 0.235 | -0.262 | 0.203 | -0.052 | 0.042 |

The first principal component has everyone pulling in the same direction. PC2 sees Clusters 2a and 2b in opposition to Cluster 3, while leaving Cluster 1 nearly unchanged. PC3 has Cluster 1 in opposition to everything else, PC4 has Clusters 2a and 3 in opposition to Cluster 2b.

If we look for the variable in each cluster that has the largest $R^2$ with the cluster as a whole (and hence may be seen as the most representative of it), for Cluster 1 that would be $D_1$ ($R^2$=0.969) followed very closely by $max|t_1|$ ($R^2$=0.963). For Cluster 2a we have the revised Matito diagnostic $revI_{ND,2021}$ ($R^2$=0.952) and $2\ln(A)/N_{val}$ ($R^2$=0.954) in very close competition, while $D_2$ is most representative of Cluster 2b ($R^2$=0.933). For Cluster 3, %TAE$_{corr}$[TPSS] and TAE$_{corr}$[CCSD(T)] both have $R^2$=0.983, but %TAE[(T)], %TAE$_X$[TPSS@HF – HF], and $A_{100}$[TPSS] all have very high $R^2$=0.958, 0.957, and 0.979, respectively.

If we introduce additional variables, such as the inverse HOMO-LUMO gap; the Bartlett 2020 diagnostic;[34] %TAE[BD(T) – CCSD(T)] or the difference between Brueckner doubles and CCSD(T); and the frontier orbital density FOD(TPSS);[35] then the separation between clusters 2a and 2b becomes blurred and the eigenvalue ratio $\lambda$(PC5)/$\lambda$(PC4) becomes much less favorable. The Bartlett diagnostic most resembles cluster 2a, $1/(\varepsilon_{LUMO}-\varepsilon_{HOMO})$ straddles the line between 2a and 2b. Cluster 2a is best looked upon as "entanglement" diagnostics; 2b by strong doubles. Cluster 3 is best regarded as pragmatic, energy-based diagnostics.

The percentage connected quadruple excitations, %TAE[$T_4$] can be taken as a pessimistic estimate for the importance of post-CCSD(T) contributions: typically, higher-order triple excitations partly compensate.[4–7] It was first pointed out by Karton et al.[32] that %TAE[(T)] has a very strong correlation with %TAE[$T_4$]: this is also seen in Table 2 ($R^2$=0.93). The percentage of correlation at the CCSD(T) level has even better correlations with $A_{100}$[TPSS], $R^2$=0.97, and %TAE$_X$[TPSS@HF – HF], $R^2$=0.96. Between these latter two $R^2$=0.99, effectively indicating that %TAE$_X$[TPSS@HF – HF] and $A_{100}$[TPSS] contain nearly the same information. Thus, the slope of the atomization energy with respect to the fraction of HF exchange is effectively driven by the difference between the HF and DFT exchange energies for the same density.

%TAE$_x$[TPSS@HF – HF] is only a fair predictor for %TAE[$T_4$] though. We also note in passing that %TAE$_{corr}$[TPSS] contains essentially the same information as %TAE$_{corr}$[CCSD(T)], at much lower computational cost.

Finally, %TAEx[TPSS – TPSS@HF], the difference between self-consistent TPSS exchange and "density-corrected" HF-TPSS, does belong in cluster 3 but clearly has weaker squared correlations with the other members of the cluster than %TAEx[TPSS@HF – HF]. Between each other, these exchange-derived diagnostics have $R^2$=0.90.

**TABLE 2.** Squared correlation matrix of the different diagnostics for the 160 closed-shell molecules in the W4-17 dataset. Heatmapping is by magnitude: deep green represents $R^2 \geq 0.96$, white $R^2 \leq 0.50$.

| $R^2$ matrix | $T_1$ | $D_1$ | max\|$T_1$\| | Truhlar M | $D_2$ | max\|$T_2$\| | $-T_eSCF$ | $2\ln(A)/N_{val}$ | $I_{ND}/(I_{ND}+I_D)$ | rev$I_{ND,2021}$ | $S_{norm}$ | %TAE[(T)] | %TAE[$T_4$] | %TAE$_{corr}$[CCSD(T)] | %TAE$_{corr}$[TPSS] | $A_{100}$[TPSS] | %TAE$_X$[TPSS@HF - HF] | %TAE$_X$[TPSS-TPSS@HF] |
|---|---|---|---|---|---|---|---|---|---|---|---|---|---|---|---|---|---|---|
| $T_1$ | 1.00 | 0.87 | 0.85 | 0.46 | 0.39 | 0.34 | 0.26 | 0.40 | 0.42 | 0.24 | 0.13 | 0.41 | 0.39 | 0.26 | 0.26 | 0.28 | 0.21 | 0.17 |
| $D_1$ | 0.87 | 1.00 | 0.92 | 0.40 | 0.31 | 0.28 | 0.26 | 0.19 | 0.25 | 0.10 | 0.03 | 0.44 | 0.39 | 0.29 | 0.30 | 0.31 | 0.24 | 0.20 |
| max\|$T_1$\| | 0.85 | 0.92 | 1.00 | 0.39 | 0.30 | 0.31 | 0.25 | 0.22 | 0.27 | 0.12 | 0.05 | 0.39 | 0.36 | 0.23 | 0.24 | 0.25 | 0.19 | 0.16 |
| Truhlar M | 0.46 | 0.40 | 0.39 | 1.00 | 0.84 | 0.73 | 0.63 | 0.56 | 0.71 | 0.48 | 0.39 | 0.40 | 0.48 | 0.28 | 0.29 | 0.27 | 0.23 | 0.29 |
| $D_2$ | 0.39 | 0.31 | 0.30 | 0.84 | 1.00 | 0.74 | 0.74 | 0.59 | 0.72 | 0.51 | 0.46 | 0.29 | 0.35 | 0.19 | 0.19 | 0.18 | 0.16 | 0.18 |
| max\|$T_2$\| | 0.34 | 0.28 | 0.31 | 0.73 | 0.74 | 1.00 | 0.66 | 0.40 | 0.49 | 0.27 | 0.23 | 0.26 | 0.34 | 0.19 | 0.19 | 0.21 | 0.20 | 0.23 |
| $-T_eSCF$ | 0.26 | 0.26 | 0.25 | 0.63 | 0.74 | 0.66 | 1.00 | 0.29 | 0.45 | 0.24 | 0.25 | 0.23 | 0.26 | 0.18 | 0.19 | 0.18 | 0.17 | 0.23 |
| $2\ln(A)/N_{val}$ | 0.40 | 0.19 | 0.22 | 0.56 | 0.59 | 0.40 | 0.29 | 1.00 | 0.82 | 0.88 | 0.84 | 0.10 | 0.15 | 0.04 | 0.04 | 0.03 | 0.02 | 0.02 |
| $I_{ND}/(I_{ND}+I_D)$ | 0.42 | 0.25 | 0.27 | 0.71 | 0.72 | 0.49 | 0.45 | 0.82 | 1.00 | 0.77 | 0.69 | 0.25 | 0.30 | 0.13 | 0.14 | 0.12 | 0.09 | 0.10 |
| rev$I_{ND,2021}$ | 0.24 | 0.10 | 0.12 | 0.48 | 0.51 | 0.27 | 0.24 | 0.88 | 0.77 | 1.00 | 0.88 | 0.05 | 0.09 | 0.01 | 0.01 | 0.00 | 0.00 | 0.00 |
| $S_{norm}$ | 0.13 | 0.03 | 0.05 | 0.39 | 0.46 | 0.23 | 0.25 | 0.84 | 0.69 | 0.88 | 1.00 | 0.01 | 0.05 | 0.00 | 0.00 | 0.00 | 0.01 | 0.00 |
| %TAE[(T)] | 0.41 | 0.44 | 0.39 | 0.40 | 0.29 | 0.26 | 0.23 | 0.10 | 0.25 | 0.05 | 0.01 | 1.00 | 0.93 | 0.91 | 0.92 | 0.90 | 0.85 | 0.77 |
| %TAE[$T_4$] | 0.39 | 0.39 | 0.36 | 0.48 | 0.35 | 0.34 | 0.26 | 0.15 | 0.30 | 0.09 | 0.05 | 0.93 | 1.00 | 0.83 | 0.83 | 0.81 | 0.78 | 0.74 |
| %TAE$_{corr}$[CCSD(T)] | 0.26 | 0.29 | 0.23 | 0.28 | 0.19 | 0.19 | 0.18 | 0.04 | 0.13 | 0.01 | 0.00 | 0.91 | 0.83 | 1.00 | 1.00 | 0.97 | 0.96 | 0.88 |
| %TAE$_{corr}$[TPSS] | 0.26 | 0.30 | 0.24 | 0.29 | 0.19 | 0.19 | 0.19 | 0.04 | 0.14 | 0.01 | 0.00 | 0.92 | 0.83 | 1.00 | 1.00 | 0.97 | 0.94 | 0.87 |
| $A_{100}$[TPSS] | 0.28 | 0.31 | 0.25 | 0.27 | 0.18 | 0.21 | 0.18 | 0.03 | 0.12 | 0.00 | 0.00 | 0.90 | 0.81 | 0.97 | 0.97 | 1.00 | 0.99 | 0.88 |
| %TAE$_X$[TPSS@HF - HF] | 0.21 | 0.24 | 0.19 | 0.23 | 0.16 | 0.20 | 0.17 | 0.02 | 0.09 | 0.00 | 0.01 | 0.85 | 0.78 | 0.96 | 0.94 | 0.99 | 1.00 | 0.90 |
| %TAE$_X$[TPSS-TPSS@HF] | 0.17 | 0.20 | 0.16 | 0.29 | 0.18 | 0.23 | 0.23 | 0.02 | 0.10 | 0.00 | 0.00 | 0.77 | 0.74 | 0.88 | 0.87 | 0.88 | 0.90 | 1.00 |

## CONCLUSION

We propose here a DFT-based diagnostic for static correlation %TAE$_X$[TPSS@HF – HF] which effectively measures how different the DFT and HF exchange energies for a given HF density are. This and %TAE$_{corr}$[TPSS] are two cost-effective a priori estimates for the adequacy of the importance of static correlation. %TAE$_X$[TPSS@HF – HF] contains nearly the same information as the earlier A diagnostic, but may be more intuitive to understand.

Principal component and variable clustering analyses of a large number of static correlation diagnostics reveal that much of the variation is explained by just two components, and almost all of it by four; these are blocked by four variable clusters (single excitations; correlation entropy; double excitations; pragmatic energetics).

## ACKNOWLEDGMENTS

This research was supported by the Israel Science Foundation (grant 1969/20) and by the Minerva Foundation (grant 2020/05). G.S. acknowledges a fellowship from the Feinberg Graduate School (Weizmann Institute). The work of E.S. on this scientific paper was supported by the Onassis Foundation—Scholarship ID: FZP 052-2/2021-2022.

# REFERENCES

[1] I. Shavitt and R.J. Bartlett, *Many – Body Methods in Chemistry and Physics* (Cambridge University Press, Cambridge, 2009).
[2] K. Raghavachari, G.W. Trucks, J.A. Pople, and M. Head-Gordon, Chem. Phys. Lett. **157**, 479 (1989).
[3] J.D. Watts, J. Gauss, and R.J. Bartlett, J. Chem. Phys. **98**, 8718 (1993).
[4] J.F. Stanton, Chem. Phys. Lett. **281**, 130 (1997).
[5] K.L. Bak, P. Jørgensen, J. Olsen, T. Helgaker, and J. Gauss, Chem. Phys. Lett. **317**, 116 (2000).
[6] A.D. Boese, M. Oren, O. Atasoylu, J.M.L. Martin, M. Kállay, and J. Gauss, J. Chem. Phys. **120**, 4129 (2004).
[7] A. Karton, P.R. Taylor, and J.M.L. Martin, J. Chem. Phys. **127**, 064104 (2007).
[8] U.R. Fogueri, S. Kozuch, A. Karton, and J.M.L. Martin, Theor. Chem. Acc. **132**, 1291 (2012).
[9] J.P. Perdew, K. Burke, and M. Ernzerhof, Phys. Rev. Lett. **77**, 3865 (1996).
[10] N.C. Handy and A.J. Cohen, Mol. Phys. **99**, 403 (2001).
[11] O. V. Gritsenko, P.R.T. Schipper, and E.J. Baerends, J. Chem. Phys. **107**, 5007 (1997).
[12] J.C. Slater, Phys. Rev. **91**, 528 (1953).
[13] J.W. Hollett and P.M.W. Gill, J. Chem. Phys. **134**, 114111 (2011).
[14] G.E. Scuseria and T. Tsuchimochi, J. Chem. Phys. **131**, 164119 (2009).
[15] A. Wasserman, J. Nafziger, K. Jiang, M.-C. Kim, E. Sim, and K. Burke, Annu. Rev. Phys. Chem. **68**, 555 (2017).
[16] J. Tao, J.P. Perdew, V.N. Staroverov, and G.E. Scuseria, Phys. Rev. Lett. **91**, 146401 (2003).
[17] A. Karton, N. Sylvetsky, and J.M.L. Martin, J. Comput. Chem. **38**, 2063 (2017).
[18] H.-J. Werner, P.J. Knowles, F.R. Manby, J.A. Black, K. Doll, A. Heßelmann, D. Kats, A. Köhn, T. Korona, D.A. Kreplin, Q. Ma, T.F. Miller, A. Mitrushchenkov, K.A. Peterson, I. Polyak, G. Rauhut, and M. Sibaev, J. Chem. Phys. **152**, 144107 (2020).
[19] D.E. Woon and T.H. Dunning, J. Chem. Phys. **98**, 1358 (1993).
[20] F. Neese, F. Wennmohs, U. Becker, and C. Riplinger, J. Chem. Phys. **152**, 224108 (2020).
[21] T.J. Lee and P.R. Taylor, Int. J. Quantum Chem. **36**, 199 (1989).
[22] C.L. Janssen and I.M.B. Nielsen, Chem. Phys. Lett. **290**, 423 (1998).
[23] I.M.B. Nielsen and C.L. Janssen, Chem. Phys. Lett. **310**, 568 (1999).
[24] E. Ramos-Cordoba, P. Salvador, and E. Matito, Phys. Chem. Chem. Phys. **18**, 24015 (2016).
[25] M.K. Kesharwani, N. Sylvetsky, A. Köhn, D.P. Tew, and J.M.L. Martin, J. Chem. Phys. **149**, 154109 (2018).
[26] X. Xu, S. Sitkiewicz, E. Ramos-Cordoba, X. Lopez, and E. Matito, in *Math/Chem/Comp 2021 - 32nd MC2 Conf. Inter Univ. Cent. Dubrovnik, 7-11 June, 2021* (2021), p. 21.
[27] P. Zanardi, Phys. Rev. A **65**, 1 (2002).
[28] P. Ziesche, O. Gunnarsson, W. John, and H. Beck, Phys. Rev. B **55**, 10270 (1997).
[29] Z. Huang, H. Wang, and S. Kais, J. Mod. Opt. **53**, 2543 (2006).
[30] H. Wang and S. Kais, Isr. J. Chem. **47**, 59 (2007).
[31] A. Peres, in *Quantum Theory: Concepts and Methods* (Springer Netherlands, Dordrecht, 2002), pp. 260–297.
[32] A. Karton, E. Rabinovich, J.M.L. Martin, and B. Ruscic, J. Chem. Phys. **125**, 144108 (2006).
[33] J. Tao, J. Perdew, V. Staroverov, and G. Scuseria, Phys. Rev. Lett. **91**, 146401 (2003).
[34] R.J. Bartlett, Y.C. Park, N.P. Bauman, A. Melnichuk, D. Ranasinghe, M. Ravi, and A. Perera, J. Chem. Phys. **153**, 234103 (2020).
[35] C.A. Bauer, A. Hansen, and S. Grimme, Chem. - A Eur. J. **23**, 6150 (2017).